\documentclass [12pt,a4paper]{article}
\usepackage{amssymb, theorem}
\usepackage{graphicx}
\usepackage{graphics}
\newtheorem{Thm}{Theorem}
\newtheorem{lemma}{Lemma}
\theorembodyfont{\rm}
\newtheorem{pl}{Example}

\def\proof{{\it Proof: }}

\def\qed{\nobreak\hfill $\square$}
\def\<{\langle}
\def\>{\rangle}

\def\sig{\sigma}
\def\fel{\textstyle{1 \over 2}}
\def\iH{{\cal H}}

\def\iA{{\cal A}}
\def\iB{{\cal B}}
\def\iC{{\cal C}}
\def\iD{{\cal D}}

\def\iM{{\cal M}}

\def\iF{{\cal F}}
\def\dim{{\rm dim}}
\def\im{\mathrm{i}}
\def\ot{\otimes}

\def\bbbr{{\mathbb R}}
\def\bbbc{{\mathbb C}}

\def\Tr{\mathrm{Tr}\,}
\def\pont{{\, \cdot \,}}

\def\tldim{{\rm \underline{dim}}}

\begin{document}
 \ \vskip 1cm 
\centerline{\LARGE {\bf Complementarity and the algebraic structure}}
\bigskip
\centerline{\LARGE {\bf of 4-level quantum systems}}
\bigskip\bigskip
\centerline{\large D\'enes Petz\footnote{E-mail: petz@math.bme.hu.
Partially supported by the Hungarian Research Grant OTKA  T068258.}$^{,4,5}$, 
Andr\'as Sz\'ant\'o\footnote{E-mail: szbandi@math.bme.hu.
Partially supported by the Hungarian Research Grant OTKA TS-49835.}$^{,5}$
and Mih\'aly Weiner\footnote{E-mail: mweiner@renyi.hu.}$^{,4}$}
\bigskip
\begin{center}
$^4$ Alfr\'ed R\'enyi Institute of Mathematics, \\H-1364 Budapest,
POB 127, Hungary
\end{center}
\medskip
\centerline{$^5$ Department for Mathematical Analysis, BUTE,}
\centerline{H-1521 Budapest, POB 91, Hungary}
\bigskip
\bigskip\bigskip
\begin{abstract}
The history of complementary observables and mutual unbiased bases
is reviewed. A characterization is given in terms of conditional
entropy of subalgebras. The concept of complementarity is extended
to non-commutative subalgebras. Complementary decompositions of a 
4-level quantum system are described and a characterization of the 
Bell basis is obtained.
\bigskip


\end{abstract}
\bigskip\bigskip

The origin of complementarity is related to the non-commutativity 
of operators describing observables in quantum mechanics. Although
the concept was born together with quantum mechanics itself, the
rigorous definition was given much later. Complementary bases or 
complementary measurements give maximal information about the quantum 
system. Complementarity is used, for example, in state estimation 
\cite{PHSz, WF} and in quantum cryptography \cite{Bruss}. When 
non-classical, say quantum, information is considered, then 
non-commutative subalgebras or subsystems of the total system should 
be chosen. The study of complementary non-commutative subalgebras is 
rather recent \cite{PDcomp}. 

In general, the knowledge of the probability distribution of 
a single physical quantity is not sufficient for determining the state 
of a system. On the other hand, a part of the information coming from  
the distributions of several quantities may be redundant. 
Intuitively, two quantities are complementary if the knowledge of 
their distributions is the most informative; i.e.  as little 
redundant as possible.

Maximal precision measurements are related to maximal Abelian 
subalgebras. However, one is also motivated to study complementarity 
for non-Abelian subalgebras. For example, units that can be considered 
in a quantum computer to be qubits are described by subalgebras that 
are isomorphic to the algebra of $2 \times 2$ matrices. One might 
be interested to choose a 
collection of qubits that are as little redundant as possible.
Conditional (or relative) entropy of subalgebras give also
some justification of the intuitive meaning of complementarity. We 
shall show that if the subalgebra $\iA$ is homogeneous and Abelian, 
then the conditional
entropy $H(\iA|\iB)$ is maximal if and only if $\iA$ and 
$\iB$ are complementary. It is shown that in general (that is, not 
assuming $\iA$ to be Abelian) complementarity cannot be characterized 
by the maximality of the relative entropy. Nevertheless, we shall also 
discuss in what sense this result supports the intuitive meaning of 
complementarity in the non-commutative case, too.

The paper contains a detailed analysis the complementary subsystems 
of two qubits. It is not surprising that the Bell basis can be 
characterized by complementarity with respect to both qubits. The 
conditional (or relative) entropy of subalgebras is not 
really computable, but for maximal Abelian subalgebras the maximum value 
of the conditional entropy is equivalent to complementarity when the 
state is tracial. The conditional entropy is estimated very concretely
in a particular example.

The content of the paper is arranged in the following way.
Section 1 is devoted to some parts of the history of the concept of 
complementarity. A complete description should be rather hard and
would require much more space. Section 2 contains the rigorous 
definitions in an algebraic setting. In the paper only the finite 
dimensional situation is discussed. Section 3 is about the relation
to the conditional entropy of subalgebras introduced long time ago
by Connes and St{\o}rmer. The main result says that complementarity is
the maximality of the conditional entropy. Section 4 and Section 5 
contains the analysis of 4-level quantum systems, or 2 qubits. This is the 
simplest framework for entanglement and actually maximal entanglement
is a kind of complementarity. It is evident that the Bell bases induces
a maximal Abelian subalgebra which is complementary to both qubits.
It turns out that the converse of this statement is true. Recently a  
conjecture appeared about the conditional  entropy of maximal Abelian 
subalgebras. In the Appendix we shall show by example that it is false. 
  
\section{A historical introduction to complementarity}

Complementarity appeared in the history of quantum mechanics in the early 
days of the theory. According to {\it Wolfgang Pauli}, the new quantum 
theory could have been called the theory of complementarity \cite{Pauli}. 
This fact shows the central importance of the notion of {\bf complementarity} 
in the foundations of quantum mechanics. Unfortunately, the  
importance did not make clear what the concept really means. The idea of 
complementarity was in connection with {\bf uncertainty relation} and 
{\bf measurement} limitations. Wolfgang Pauli wrote to Heisenberg 
in 1926: ``{\it One may view the world with the $p$-eye and one may view 
it with the $q$-eye but if one opens both eyes simultaneously then one 
gets crazy}''. The distinction between {\bf incompatible} and 
{\bf complementary} observables was not really discussed. This can be
the reason that ``complementarity'' was avoided in the book \cite{vN} of 
von Neumann, although the mathematical foundations of quantum theory were 
developed in a generally accepted way. The concept of complementarity was 
not clarified for many years, but it was accepted  that the pair of 
observables of {\bf position} and {\bf momentum} must be a typical and 
important example (when complementarity means a relation of observables). 

The canonically conjugate position and momentum, $Q$ and $P$,
are basic observables satisfying the {\bf commutation relation},
$$
(QP - PQ)f =\im f \qquad (f \in \iD)
$$
which holds on a dense domain $\iD$ (for example, on the Schwartz functions
in $L^2(\bbbr)$). The {\bf uncertainty relation},
$$
\Delta (Q,f )\,\Delta (P,f ) \ge \frac 12  \qquad (f \in \iD)
$$
holds on the same domain. (Recall that $\Delta (A,f )=\sqrt{\<f, A^2f\>
-\<f,Af\>^2}$ is the variance of the observable $A$ in the vector state $f$.)

The {\bf Fourier connection} $P = \iF^{-1}Q\iF$  extends also to the
spectral measures $E^P(\pont)$ and $E^Q(\pont)$, so that one has 
$$
E^P(H) = \iF^{-1}E^Q(H)\iF
$$
for all Borel sets $H\subset \bbbr$. From the Fourier relation one can
deduce that  $E^Q(H_1)f = f$ and $E^P(H_2)f = f$ for some
vector $f$ and bounded sets $H_1,H_2 \subset \bbbr$ may hold only
in the trivial case $f=0$. Therefore, the following well-known
relations for the spectral projections are obtained:
$$
E^Q(H_1)\land E^P(H_2)=E^Q(H_1)\land E^P(\bbbr\setminus H_2)=
E^Q(\bbbr\setminus H_1)\land E^P(H_2)=0
$$
for all bounded $H_1, H_2\subset \bbbr$, where $\land$ denotes the 
greatest lower bound in the lattice of projections. For some people, these 
relations show the {\bf complementarity} of  $Q$ and $P$.
(It may be of interest to note that $E^Q(\bbbr\setminus H_1)\land
E^P(\bbbr\setminus H_2)\ne O$, for all bounded $H_1$ and $H_2$
\cite{Len}.) $Q$ and $P$ are {\bf totally non-commutative}: There 
are no vectors with respect to which $Q$ and $P$ commute. People have
agreed that position  and momentum are complementary observables. This 
opinion was supported by the fact that two observables cannot be 
measured or tested together \cite{B-L}.

If the Hilbert space of the quantum system is finite dimensional, then
the total non-commutativity of two observables is typical. If 
complementarity means maximal incompatibility, then the definition
must be different.

{\it Herman Weyl} used the finite Fourier transform to approximate the
relation of $P$ and $Q$ in finite dimensional Hilbert spaces \cite{Weyl}. 
Let $|0\>,|1\>,\dots ,|n-1\>$ be an orthonormal basis in an $n$-dimensional
Hilbert space. The transformation
\begin{equation}\label{F:Fourier}
V_n: |i\> \mapsto \frac{1}{\sqrt{n}}\sum_{j=0}^{n-1} \omega^{ij} |j\> 
\qquad (\omega=e^{2\pi \im /n})
\end{equation}
is a unitary and it is nowadays called {\bf quantum Fourier transform}. 
If the operator $A=\sum_i \lambda_i |e_i\>\<e_i|$ is diagonal in the given
basis and $B=V_n^*AV_n$, then the pair $(A,B)$ approximates $(Q,P)$ when 
the eigenvalues are chosen properly. 
 
The complementarity of observables of a finite quantum system was emphasized
by Accardi in 1983 during the Villa Mondragone conference \cite{Ac}. His
approach is based on conditional probabilities. If an observable is measured
on a copy of a quantum system and another observables is measured on
another copy (prepared in the same state), then one measurement does not
help to guess the outcome of the other measurement, if all conditional
probabilities are the same. If the eigenvectors of the first observable are
$\xi_i$'s, the eigenvectors of the second one are $\eta_j$'s and the
dimension of the Hilbert space is $n$, then complementarity means
\begin{equation}\label{E:compi}
|\<\xi_i,\eta_j\>|=\frac{1}{\sqrt{n}}.
\end{equation}
It is clear that the complementarity of two observables is actually the 
property of the two eigenbases, so it is better to speak about complementary
bases. The Fourier transform (\ref{F:Fourier}) moves the standard basis
$|0\>,|1\>,\dots ,|n-1\>$ to a complementary basis  $V_n|0\>,V_n|1\>,
\allowbreak \dots , V_n|n-1\>$. The complementarity (\ref{E:compi}) is
often called {\bf value complementarity} and it was an important subject in
the work of Schwinger \cite{Redei, Sch}.

In connection with complementarity, Kraus made a conjecture 
about the entropy of two observables \cite{Kr} which was proved by
Maasen and Uffink \cite{M-U}.

The goal of state determination is to recover the state of a quantum system 
by measurements. If the Hilbert space is $n$ dimensional, then the
density matrix of a state contains $n^2-1$ real parameters. If a measurement
is repeated on many copies of the same system, then $n-1$ parameters can 
be estimated. Therefore, at least $n+1$ different measurement should be
performed to estimate the $n^2-1$ parameters. A measurement can be identified
with a basis. Wootters and Fields argued that in the optimal situation 
estimation scheme the $n+1$ bases must be pairwise complementary 
\cite{WF}. Instead of pairwise complementary bases, Wootters and Fields used 
the expression ``{\bf mutually unbiased bases}'' and this terminology has 
become popular. A different kind of optimality of the complementary bases
was obtained in \cite{PHM} in terms of the determinant of the average mean 
quadratic error matrix.

While Kraus was interested in the number of non-unitarily equivalent complementary
pairs, after publication of the paper of Wootters and Fields, the maximum number
of mutually unbiased bases become a research subject. (It is still not known if
for any dimension $n$ the upper bound $n+1$ is accessible, \cite{TZ}.) 

\section{Complementary subalgebras}

Let $\iH$ be an $n$-dimensional Hilbert space with an orthonormal basis
$e_1,e_2,\dots, e_n$. A unit vector $\xi \in \iH$ is {\bf complementary}
with respect to the given basis $e_1,e_2,\dots, e_n$ if
\begin{equation}\label{E:c}
|\< \xi, e_i\>|=\frac{1}{\sqrt{n}}\qquad (1 \le i \le n).
\end{equation}

When the Hilbert space $\iH$ is a tensor product $\iH_1 \ot \iH_2$, then a unit
vector complementary to a product basis is called {\bf maximally entangled state}.
(If a vector is complementary to a product basis, then it is complementary to
any other product basis.) When $\dim \iH_1=\dim \iH_2=2$, then the {\bf Bell basis}
consists of maximally entangled states.

(\ref{E:c}) is equivalent to the formulation that the vector state $|\xi\>\<\xi|$
gives the uniform distribution when the measurement $|e_1\>\<e_1|, \dots, 
|e_n\>\<e_n|$ is performed:
$$
\Tr |\xi\>|\< \xi|\,| e_i\>\<e_i| =\frac{1}{n}\qquad (1 \le i \le n).
$$
The unital subalgebra generated by $|\xi\>|\< \xi|$ consists of operators
$\lambda |\xi\>|\< \xi|+\mu |\xi\>|\< \xi|^\perp$ ($\lambda, \mu \in \bbbc$), 
while the algebra generated by the orthogonal projections $| e_i\>\<e_i|$ is 
$\{\sum_i \lambda_i | e_i\>\<e_i|:\lambda_i \in \bbbc\}$. Relation (\ref{E:c})
can be reformulated in terms of these generated subalgebras.

\begin{Thm}\label{T:uj}
Let $\iA_1$ and $\iA_2$ be subalgebras of $M_k(\bbbc)$ and let $\tau:=\Tr /k$
be the normalized trace. Then the following conditions are equivalent:
\begin{enumerate}
\item[(i)] 
If $P \in \iA_1$ and $Q \in \iA_2$ are minimal projections,
then $\tau (PQ)=\tau(P)\tau(Q)$.
\item [(ii)]
The subalgebras $\iA_1$ and  $\iA_2$ are quasi-orthogonal in $M_n(\bbbc)$, that
is the subspaces $\iA_1 \ominus \bbbc I$ and  $\iA_2 \ominus \bbbc I$ are orthogonal.
\item [(iii)]  $\tau(A_1 A_2)=\tau(A_1)\tau(A_2)$ if $A_1 \in \iA_1$,  $A_2 \in 
\iA_2$.
\item [(iv)]  If $E_1:\iA \to \iA_1$ is the trace preserving conditional expectation, then
$E_1$ restricted to $\iA_2$ is a linear functional (times $I$). 
\end{enumerate}
\end{Thm}

This theorem was formulated in \cite{PDcomp} and led to the concept of complementary
subalgebras. Namely $\iA_1$ and $\iA_2$ are complementary if the conditions of the 
theorem hold. As we explained above complementary maximal Abelian subalgebras
is a popular subject in the form of the corresponding bases. We note that
complementary MASA's was studied also in von Neumann algebras \cite{Popa}

Two orthonormal bases are connected by a unitary. It is quite obvious that two bases
are mutually unbiased if and only if the absolute value of the elements of the 
transforming unitary is the same, $1/\sqrt{n}$ when $n$ is the dimension. This
implies that construction of mutually unbiased bases is strongly related (or equivalent)
to the search for Hadamard matrices. 

Let $\iA_1$ and $\iA_2$ be subalgebras of $M_k(\bbbc)$ and assume that both 
subalgebras are isomorphic to $M_m(\bbbc)$. Then $k=mn$ and we can assume that
$\iA_1=\bbbc  I_n\ot M_m(\bbbc)$. There exists a unitary $W$ such that $W\iA_1 W^*=
\iA_2$. The next theorem characterizes $W$ when $\iA_1$ and $\iA_2$ are 
complementary \cite{OPSz, PDcomp}. (On the matrices the Hilbert-Schmidt inner 
product $\< A,B\>=\Tr A^*B$ is considered.)

\begin{Thm}\label{T:1}
Let $E_{i}$ be an orthonormal  basis in $M_n(\bbbc)$ and let $W= 
\sum_{i} E_{i} \ot W_{i} \in M_n(\bbbc) \ot M_m(\bbbc)$ be a unitary. 
The subalgebra $W(\bbbc  I_n\ot M_m(\bbbc))W^*$ is complementary 
to $\bbbc  I\ot M_m(\bbbc)$ if and only if 
$$
\frac{m}{n} \sum_{k} |W_{k}\> \<W_{k}|
$$
is the identity mapping on $M_m(\bbbc)$. 
\end{Thm}

The condition in the Theorem cannot hold if $m<n$ and in the case $n=m$ 
the condition means that  $\{W_{k}:1 \le k \le n^2\}$ is an orthonormal basis 
in $ M_m(\bbbc)$.

A different method for the construction of complementary subalgebras is indicated 
in the next example.

\begin{pl}\label{pl:Sch}
Assume that $p>2$ is prime. Let  $e_0, e_1,\dots, e_{p-1}$ be a basis 
and let $X$ be the unitary operator permuting the basis vectors cyclically:
$$
Xe_i=\left\{ \begin{array}{ll}
e_{i+1}\quad & \hbox{if } 0 \le i \le n-2,
\\
e_0 &\hbox{if } i=n-1.\end{array} \right.
$$
Let $q:=e^{\im 2\pi/p}$ and define another unitary by $Ze_i=q^i e_i$.
Their matrices are as follows.
 \[
X=
\left[\begin{array}{ccccc}
0 & 0 &  \cdots & 0 & 1 \\
1 & 0 &  \cdots & 0 &0 \\
0 & 1 &  \cdots & 0 &0\\
\vdots & \vdots & \ddots &\vdots & \vdots \\
0 & 0  & \cdots &1 & 0
\end{array}\right], \qquad
Z=
\left[\begin{array}{ccccc}
1 & 0 &0& \cdots & 0 \\
0 & q &0 & \cdots & 0 \\
0 & 0& q^2  & \cdots & 0 \\
\vdots & \vdots &\vdots & \ddots & \vdots \\
0 & 0 &0&  \cdots & q^{p-1}
\end{array}\right].
\]
 
It is easy to check that $ZX=qXZ$ or more generally the relation
\begin{equation}\label{eq4.1}
(X^{k_1} Z^{\ell_1})( X^{k_2} Z^{\ell_2}) = 
q^{k_2 \ell_1 } X^{k_1 + k_2} Z^{\ell_1 + \ell_2}.
\end{equation}
is satisfied. The unitaries
$$
\{X^j Z^k\, :\, 0 \le j,k \le p-1\}
$$
are pairwise orthogonal. 

For $0 \le k_1,\ell_1,k_2,\ell_2 \le p-1$ set
\[
\pi (k_1,l_1,k_2,l_2) = X^{k_1}Z^{\ell_1} \otimes X^{k_2} Z^{\ell_2}.
\]

From (\ref{eq4.1}) we can compute
\begin{equation}\label{eq4.3}
\pi(u) \pi(u') = q^{-u\circ u'} \pi(u')\pi(u),
\end{equation}
where
\[
u \circ u' = k_1\ell_1' - k_1' \ell_1  + k_2 \ell_2' - k_2' \ell_2  
\qquad ({\rm mod} \,\,p)
\]
for $u = (k_1,\ell_1,k_2,\ell_2)$ and $u'= (k_1',\ell_1',k_2',\ell_2')$. 
Hence $\pi(u) $ and $\pi(u')$ commute if and only if $u \circ u'$ equals zero.

We want to define a homomorphism $\rho:M_p(\bbbc) \to M_p(\bbbc)\ot M_p(\bbbc)$
such that
$$
\rho(X)=\pi (k_1,l_1,k_2,l_2)\quad \mbox{and}\quad
\rho(Z^{u\circ u'}) = \pi(u')
$$
when $u \circ u'\ne 0$. Since the commutation relation (\ref{eq4.3}) is the same  
as that for $X$ and $Z^{u\circ u'}$, $\rho$ can be extended to an embedding 
of $M_p(\bbbc)$ into $M_p(\bbbc)\ot M_p(\bbbc)$. Let $\iA(u,u') \subset 
M_p(\bbbc)\ot M_p(\bbbc)$ be the range. This is a method to construct subalgebras.
For example, if
$$
\pi(u)=X\ot X \quad \mbox{and}\quad \pi(u')= Z \ot Z,
$$
then the generated subalgebra $\iA(u,u')$ is obviously complementary to
$\bbbc I \ot M_p(\bbbc)$ and $M_p(\bbbc) \ot \bbbc I$. (At this point we
used the condition $p>2$, since this implies that $X$ and $Z$ do not
commute.) \qed
\end{pl}
 
The idea of the above example is used by Ohno to construct $p^2+1$
complementary subalgebras in $M_p(\bbbc)\ot M_p(\bbbc)$ \cite{Ohno}.

\section{Conditional entropy}

Let $\iA$ and $\iB$ be subalgebras of $\iM\equiv M_n(\bbbc)$.
For a state $\psi$ on $\iM$ the {\bf conditional entropy} of the 
algebras $\iA$ and $\iB$ is defined as
\begin{equation}\label{E:cond}
H_\psi (\iA | \iB):= \sup\Big\{ \sum_i \lambda_i \bigg(S(\psi_i|_\iA \, 
\|\, \psi|_\iA) - S(\psi_i|_\iB \, \|\, \psi|_\iB) \bigg)\Big\}\,
\end{equation}
where the supremum is taken over all possible decomposition of $\psi$ 
into a convex combination $\psi=\sum_i \lambda_i \psi_i$ of states
and $S(\pont||\pont)$ stands for the relative entropy of states. This
concept was introduced by Connes and St{\o}rmer in 1975 \cite{CS}
and was called relative entropy of subalgebras. Since in the case of
commutative algebras, the quantity becomes the usual conditional entropy, 
see Chap. 10 in \cite{Stormer}, we are convinced that conditional entropy 
is the proper terminology.

If $\iB=\bbbc I$, then  $H_\psi (\iA | \iB)$ is the entropy 
$$
H_\psi (\iA)=\sup\Big\{ \sum_i \lambda_i \bigg(S(\psi_i|_\iA \, 
\|\, \psi|_\iA)\bigg)\Big\}
$$ 
of the subalgebra $\iA$ \cite{OP}. The quantity $H_\psi (\iA)$ is 
heuristically the amount of information contained in the subalgebra 
$\iA$ about the state. In this spirit, the conditional entropy measures 
the information difference carried by $\iA$ and $\iB$ together with 
respect to $\iB$. Formally we can state much less. For example, 
$H_\psi (\iA | \iB)=0$ if and only if $\iA \subset \iB$ \cite{Stormer}. 
We have $H_\psi (\iA | \iB) \le H_\psi (\iA )$ and  in probability theory 
the equality is equivalent to the independence of $\iA$ and $\iB$ 
(with respect to $\psi)$. Here we are interested in the tracial state 
in the role of $\psi$ and want to study the relation of the maximality of
the conditional entropy to the complementarity of the subalgebras.

One may wonder whether at taking supremum, we should really consider 
all possible decompositions. Indeed, from the point of view of 
actual calculations, it is a rather unfortunate thing, as in 
some sense they are ``too many'' to parametrize. 

\begin{lemma}
It is enough to take the supremum in (\ref{E:cond}) over all possible 
decomposition of $\psi$ into a convex combination $\psi=\sum_i \lambda_i 
\psi_i$ of linearly independent states (over $\bbbr$). 
\end{lemma}

\proof
Let $\psi=\sum_{i=1}^k \lambda_i \psi_i$ be a decomposition of $\psi$ 
into a convex combination of states. Without the loss of generality, we 
may assume that all $\lambda_i$ weights are nonzero. (States with zero 
weights can be simply left out both from the decomposition and from the 
expression of the entropy, too.) Suppose there is a nontrivial 
real-linear dependence between the states appearing in the 
decomposition; that is, we have that $\sum_{i=1}^k \alpha_i\psi_i=0$ for 
a collection of nontrivial real coefficients $\alpha_i$. 
Then with
$$
\lambda_i(t):=\lambda_i + \alpha_i t,
\;\;\;\;\textrm{we have}\;\;\;\;
\psi  = \sum_{i=1}^k \lambda_i(t)\psi_i. 
$$
Since $\min\{\lambda_i\}>0$, there is an interval $I\subset \bbbr$ having 
$0$ as an interior point, such that $\lambda_i|_I \geq 0$ for all indices. 
In fact, it is rather evident, that there exists a maximal such interval, say
$I_{\rm max}$, and that $I_{\rm max} = [a,b]$ is closed and there exist
some indices $j_a$ and $j_b$ such that at the endpoints of the interval 
$\lambda_{j_a}(a)=\lambda_{j_b}(b)=0$. Since the function $h$ defined 
by the formula 
$$
h(t):= \sum_i \lambda_i(t) \bigg(S(\psi_i|_\iA \, \|\,
\psi_i|_\iA) - S(\psi_i|_\iB \, \|\, \psi|_\iB)\bigg)
$$
is a polynomial of order at most one, we have that $h(0)\leq 
\max\{h(a),h(b)\}$. In other words, we may change the original sum 
(i.e. the sum at parameter $t=0$) by letting $t=a$ or $t=b$ in such a 
way, that its value will not decrease. However, the sum at parameter $t=a$ 
$(t=b)$ corresponds to a convex combination of the states $\psi_i$ with 
$i\neq j_a$ ($i\neq j_b$). In other words, using the linear dependence we can 
eliminate at least one of the states appearing in the decomposition in such a 
way that the sum will surely not decrease. This verifies our claim since we may
repeat this process until the set of states in question will be linearly 
independent. \qed

It is a consequence of the lemma that in the finite dimensional case, 
the word ``supremum'' appearing in the definition of conditional entropy 
can be replaced by the word ``maximum''.  (A standard argument relies on 
the continuity of the entropy functional and the compactness of the state 
space.) 

In what follows the reference state $\psi$ will be always the unique 
normalized tracial state $\tau:=\Tr /n$ on $\iM\equiv M_n(\bbbc)$. So 
we shall omit the indication of the reference state and simply write 
$H(\iA|\iB)$ instead of $H_\tau(\iA|\iB)$. Also, instead of the states 
$\psi_i$, it will be often convenient to work with their density matrices 
$\rho_i$ with respect to $\tau$. It is an easy exercise to check that the 
conditional entropy is expressed with density matrices as  
\begin{equation}\label{eta}
H(\iA | \iB) =\sup \Big\{
\sum_i \lambda_i \bigg(\tau(\eta (E_\iB \rho_i)) - \tau(\eta(E_\iA 
\rho_i)\bigg)\Big\},
\end{equation}
where $E_\iA: \iM \to \iA$ and $E_\iB: \iM \to \iB$ are
the $\tau$-preserving conditional expectations,  $\eta(t)=-t\log t$, 
and the supremum is taken over all possible convex decompositions of 
the identity $I=\sum_i \lambda_i \rho_i$.

Our primary interest concerns the case when the subalgebras in question
are either maximal Abelian or isomorphic to some full matrix algebras. 
The two cases will be discussed together; for our argument it will be 
enough to assume that all minimal projections of $\iA$ have the 
same trace. Such subalgebra $\iA$ will be called {\bf homogeneous}.
Suppose that for every minimal projection $p\in \iA$ we have $\tau(p)=d$.
Then for every density operator $\rho$ and minimal projection $p\in \iA$, 
we have that
$$
\tau(\eta(E_\iA(\rho))) \ge \tau(\eta(p/d)) = \log d,
$$
and equality holds if and only if $d E_\iA(\rho)$ is a 
a minimal projection of $\iA$, which is trivially further equivalent 
with the fact that the range of $\rho$ is contained in the range of a 
minimal projection of $\iA$. On the other hand,
$$
\tau(\eta(E_\iB(\rho))) \le \tau(\eta(I)) = 0.
$$
This implies that
\begin{equation}\label{E:bound}
H(\iA|\iB)\le -\log d\, .
\end{equation}
In general it is easy to give some sufficient conditions
ensuring that in the above inequality one has equality.
When $\iA$ is Abelian, we can also give a simple necessary condition.

\begin{lemma}\label{L:7}
Let  $\iA$ be a homogeneous subalgebra such that $\tau(p)=d$ for the minimal
projections $p \in \iA$. If there exists a decomposition $I = \sum_i 
\lambda_i p_i$ of the identity such that $\lambda_i>0$ and $p_i$ are 
minimal projections of $\iA$ satisfying $E_\iB(p_i)=d I$, then equality holds
in (\ref{E:bound}).
\end{lemma}

\proof
It is enough to give a lower estimate for the conditional entropy:
\begin{eqnarray} \nonumber
H(\iA|\iB) &\geq&
\sum_i \lambda_i d \bigg(\tau(\eta (E_\iB (p_i/d))) - 
\tau(\eta(E_\iA (p_i/d))\bigg)
\\
&=&
\sum_i \lambda_i d \bigg(\tau(\eta (I)) -
\tau(\eta(p_i/d)\bigg) = 
\sum_i \lambda_i d  \log (1/d)
= - \log d,
\end{eqnarray}
since from $1=\tau(I)=\tau(\sum_i \lambda_i p_i)$ we get that 
$\sum_i\lambda_i d =1$. \qed

\begin{Thm} Let $\iA$ and $\iB$ be subalgebras of 
$M_n(\bbbc)$. Assume that $\iA$ is Abelian and homogeneous.
Then the subalgebras $\iA$ and $\iB$ are complementary
if and only if $H(\iA|\iB)$ is maximal.
\end{Thm} 

\proof
If $\iA$ and $\iB$ are complementary, then for the minimal projections
$p_i$ of $\iA$, $\sum_i p_i=I$ and $E_\iB(p_i)=d I$ hold. So Lemma \ref{L:7}
tells us that the conditional entropy is $-\log d$. 

Assume now that $H(\iA|\iB)= -\log d$. Then there exists a decomposition 
$I= \sum_i \lambda_i \rho_i$ of the identity  into a convex 
combination of density operators such that $E_\iB(\rho_i)=I$ and $q_i:
= E_\iA(\rho_i)/n$ are minimal projections of $\iA$. 

Suppose that the image under the trace-preserving expectation $E$ onto a 
subalgebra of a positive operator $a$ is a multiple of a minimal 
projection $p$ of the subalgebra. Then $x:=(I - p)a(I -p)$ is a 
positive operator for which 
$$
E(x)= (I-p) E(a)(I-p)=0,
$$
and hence $x=0$. It follows that  $(I -p)\sqrt{a}=0$ and we conclude $pa=ap=a$.

Applying the above, we have that for every minimal projection $q$ of 
$\iA$
$$
q=
qI = q\sum_i \lambda_i \rho_i = 
q\sum_i \lambda_i q_i\rho_i = \sum_i \lambda_i qq_i \rho_i 
=\sum_{\{i: q_i=q\}} \lambda_i q_i\rho_i  
=\sum_{\{i: q_i=q\}} \lambda_i \rho_i,
$$
since the product $qq_i$ is zero, when $q_i\neq q$ and $q_i$ when 
$q_i=q$. (Note that this is the point where we have used the fact 
the $\iA$ is Abelian). As $E_\iB(\rho_i)=I$, the 
above decomposition of $q$ shows that $E_\iB(q)$ is a multiple of the 
identity, and hence (as $q$ was arbitrary, and the minimal projections 
of $\iA$ span the whole algebra $\iA$) that $\iA$ is quasi-orthogonal to 
$\iB$. \qed

Let $\iA$ and $\iB$ be subalgebras of $M_n(\bbbc)$. Assume that $\iA$ is 
Abelian and homogeneous and choose a homogeneous algebra 
$\iC$ such that $\iA$ is maximal Abelian subalgebra of $\iC$. If $\iA$ 
and $\iB$ are complementary, then $H(\iC|\iB)$ is maximal (that is, 
equals $H(\iC)$). However, $\iC$ and $\iB$  is not necessarily 
complementary,  in fact it 
is fairly easy to come up with an example in which their intersection 
is not trivial. Hence the conditional entropy cannot characterize the 
complementarity of subalgebras in the general case.

Suppose we are dealing with two subsystems (that is, subalgebras) 
$\iB,\iC$ of a finite level quantum system (that is, $M_n(\bbbc)$).
Knowing the restriction of the state to $\iB$ might help in predicting 
results of measurements performed on $\iC$. However, a (maximal 
precision) measurement on $\iC$ corresponds to a maximal Abelian 
subalgebra of $\iC$. Hence the natural interpretation of complementarity 
suggests that complementarity of $\iB$ and $\iC$ should mean 
that for {\it all} maximal Abelian subalgebras $\iA$ of $\iC$ the 
conditional entropy $H(\iA,\iB)$ should be maximal. By what was proved 
in this section, if $\iA$ is homogeneous, then this condition indeed 
characterizes complementarity.

\section{4-level quantum systems}

A 4-level quantum system is mathematically the Hilbert space $\bbbc^4$ 
or the algebra $\iM:=M_4(\bbbc)$. We are interested in two kinds of 
subalgebras. 

An {\bf F-subalgebra} is a subalgebra isomorphic to $M_2(\bbbc)$. 
``F'' is the abbreviation of "factor", the center of such a subalgebra 
is minimal, $\bbbc I$. If our 4-level quantum system is regarded as two qubits, 
then an F-subalgebra may correspond to one of the qubits. When the F-subalgebra 
$\iA_0$ describes a ``one-qubit-subsystem'', then the relative commutant 
$\iA':=\{ B\in \iM: BA=AB$ for every $A\in \iA\}$ corresponds to the other qubit. 
If $\iA$ is an F-subalgebra of $\iM$, then we may assume that $\iM= \iA \ot \iA'$.
 
An {\bf M-subalgebra} is a  maximal Abelian subalgebra, 
equivalently, it is isomorphic to $\bbbc^4$. (M is an abbreviation of 
``MASA'', the center is maximal, it is the whole subalgebra.) An M-subalgebra 
is in relation to a {\bf von Neumann measurement}, its minimal projections give a 
partition of unity.

Both the F-subalgebras and the M-subalgebras are 4 dimensional. We define a 
{\bf P-unitary} as a self-adjoint traceless unitary operator. The eigenvalues 
of a P-unitary from $\iM$ are $-1,-1,1,1$. An {\bf F-triplet} $(S_1,S_2,S_3)$ 
consists of P-unitaries such that $S_3=\im S_1 S_2$. An {\bf M-triplet} 
$(S_1,S_2,S_3)$ consists of P-unitaries such that $S_3=S_1 S_2$. One can see that 
if $(S_1,S_2,S_3)$ is an X-triplet, then the linear span of $I,S_1,S_2,S_3$ is 
an X-subalgebra, X=F, M.

\begin{pl}\label{Pl:Fourier}
Consider the unitary $W=V_{n^2}$ defined in (\ref{F:Fourier}) as an $n \times n$ 
block-matrix with entries from $M_n(\bbbc)$. Then the entries form an orthonormal 
basis in $M_n(\bbbc)$ and Theorem \ref{T:1} tells us that the Fourier transform 
can be used to construct a complementary pair. 

The Fourier transform sends the standard basis into a complementary one
but it can produce non-commutative complementary subalgebras as well. If
$n=2$, then we get the following two F-triplets
$$
\sigma_0 \ot \sigma_1,  \quad \sigma_0 \ot \sigma_2,  \quad
\sigma_0 \ot \sigma_3
$$
and
$$
\fel(-\sigma_2 \ot \sigma_0-\sigma_2 \ot \sigma_3+\sigma_3 \ot \sigma_0),
\fel(-\sigma_2 \ot \sigma_0-\sigma_2 \ot \sigma_3+\sigma_3 \ot \sigma_0+
\sigma_3 \ot \sigma_3),
-\sigma_1 \ot \sigma_0.
$$
\qed
\end{pl}


In the Hilbert space $\bbbc^4=\bbbc^2 \ot \bbbc^2$ the standard product basis is
$|00\>, |01\>$, $|10\>, |11\>$. The {\bf Bell basis}
$$
\frac{1}{\sqrt{2}}(|00\>+|11\>),\quad \frac{1}{\sqrt{2}}(|01\>+|10\>), \quad
\frac{1}{\sqrt{2}}(|00\>-|11\>), \quad \frac{1}{\sqrt{2}}(|01\>+|10\>)
$$
consists of maximal entangled vectors and so it is complementary to the standard 
product basis. The Bell basis has important applications, for example, the 
teleportation of a state of a qubit. We show in Theorem \ref{T:Bell} below 
that the above complementarity property characterizes the Bell basis. Up to 
local unitary transformations, the Bell basis is unique.

The operators diagonal in the Bell basis form an M-subalgebra which is generated
by the M-triplet
\begin{equation}\label{Be-tri}
(\sigma_1 \ot \sigma_1,  \quad \sigma_2 \ot \sigma_2, \quad \sigma_3 \ot \sigma_3).
\end{equation}
We call this standard {\bf Bell triplet}.

\begin{Thm}\label{T:Bell}
Let $\iA$ be an F-subalgebra of $\iM$. Assume that $(X,Y,Z)$ is an M-triplet 
which is orthogonal to $\iA$ and $\iA'$. Then there are F-triplets $(A_1,A_2,A_3) 
\in \iA$ and $(B_1,B_2,B_3) \in \iA'$ such that
$$
X=A_1B_1, \quad Y=A_2B_2, \quad Z=A_3B_3.
$$
\end{Thm}

\proof
Take an expansion
$$
X=\sum_{i=1}^3 (x_i \cdot \sigma) \ot \sigma_i,
$$
where $x_i\in \bbbr^3$. Then
\begin{eqnarray*}
X^2&=&\sum_{i,j=1}^3 (\<x_i, x_j\> \sig_0+\im (x_i \times x_j)\cdot \sigma) 
\ot \sigma_i \sigma_j 
=\sum_{i=1}^3 (\<x_i, x_i\> \sig_0+\im (x_i \times x_i)\cdot \sigma) \ot \sigma_0 \cr &&
+\sum_{i<j} (\<x_i, x_j\> \sig_0+\im (x_i \times x_j)\cdot \sigma) \ot \sigma_i \sigma_j
+\sum_{i>j} (\<x_i, x_j\> \sig_0+\im (x_i \times x_j)\cdot\sigma) \ot \sigma_i 
\sigma_j \cr
&=& \sum_{i=1}^3 \<x_i, x_i\> \sig_0 \ot  \sigma_0 +\sum_{i<j} (2\im (x_i 
\times x_j)\cdot \sigma) \ot \sigma_i \sigma_j\,.
\end{eqnarray*}
We conclude that 
$$
x_i \times x_j=0
$$ 
when  $i\ne j$. All the three vectors $x_i$ cannot be 0, so we may assume 
that $x_1 \ne 0$. Then
there are $\lambda,\mu \in \bbbr$ such that $x_2=\lambda x_1$ and 
$x_3=\mu x_1$. So $X=(x_1\cdot\sigma) \otimes (\sig_1+\lambda\sig_2+
\mu \sig_3)$  Since $\sum_{i=1}^3 \<x_i, x_i\>=1$,
for an appropriate number $\kappa$, the matrices
$A_1:=\kappa(x_1\cdot\sigma) \otimes I$ and $B_1:=\kappa^{-1}I 
\otimes (\sig_1+\lambda\sig_2+\mu \sig_3)$ are P-unitaries and the relations 
$$
X=A_1B_1, \quad A_1X=XA_1,\quad B_1X=XB_1
$$
hold.

Next we show that $A_1Y=-YA_1$. Changing the Pauli matrices by unitary 
transformation, we may assume
that $A_1=\sigma_1 \ot I$ and $B_1=I \ot \sig_1$. The commutant of 
$X=\sig_1 \ot \sig_1$ is the linear
span of the 8 matrices
$$
\{I, \sig_1\ot I, I\ot \sig_1, \sig_1 \ot \sig_1,\} \cup \{\sig_i
\ot \sig_j\}_{i,j=2}^3\,.
$$
Recall that $Y$ is orthogonal to $\iA$ and $\iA'$, so it must 
be the linear combination of the matrices
$\{\sig_i\ot \sig_j\}_{i,j=2}^3$.  All of them anticommute with 
$A_1$, therefore so does $Y$.
The matrix $A_1Y$ is the linear combination of the matrices 
$\{\sig_1\sig_i\ot \sig_j\}_{i,j=2}^3$, which
implies $A_1Y \perp \iA'$.

Since $A_1Y=-YA_1$, it follows that $\{A_1,Y,-\im A_1Y\}$ is an 
F-triplet which generates the F-subalgebra $\iA_1$. The F-subalgebras 
$\iA_1$ and $\iA'$ are complementary, hence $\iA_1'$ and $\iA$ are 
complementary as well. The intersection of $\iA_1'$ and $\iA'$ is different
from $\bbbc I $\cite{OPSz}, therefore it contains a non-trivial projection $Q$
which must have trace 2. It follows that the intersection contains a P-unitary 
$B_2:=I-2Q$. So $B_2$ commutes with $Y$ and let 
$A_2:=B_2Y=YB_2$. We check that $B_1$ and $B_2$ anticommute:
$$
B_1 B_2=(A_1X)(YA_2)=A_1YXA_2=-Y(A_1X)A_2=-(B_2 A_2)B_1A_2=-B_2B_1,
$$
since $B_1=A_1X$, $XY=YX$, $A_1Y=-YA_1$, $Y=B_2A_2$, $A_2B_1=B_1A_2$.
Similarly, $A_1$ and $A_2$ anticommute. $Y=A_2B_2$ is obvious. If
$A_3-=\im A_1A_2$ and $B_3-=\im B_1B_2$, then both $(A_1,A_2,A_3)$
and $(B_1,B_2,B_3)$ are F-triplets. Finally,
$$
Z=XY=A_1B_1A_2B_2=A_1A_2 B_1B_2=A_3 B_3
$$
and the proof is complete. \qed

If the operators $A_i$ and $B_i$ are identified with $\sigma_i$ ($i=1,2,3$)
in the theorem, then the triplet $X,Y,Z)$ can be identified with the standard 
Bell triplet (\ref{Be-tri}).

\begin{pl}\label{Pl:CAR}
Let $\iA$ be the algebra generated by the operators $a_1, a_1^*,a_2,a_2^*$
satisfying the {\bf canonical anticommutation relations}:
$$
\{a_1,a_1^*\}=\{a_2,a_2^*\}=I, 
\{a_1,a_1\}=\{a_1,a_2\}=\{a_1,a_2^*\}=\{a_2,a_2\}=0,
$$
where $\{A,B\}:=AB+BA$. Let $\iA_1$ be the subalgebra generated $a_1$ and 
$\iA_2$ be the subalgebra generated $a_2$. Then $\iA_1$ and $\iA_2$ are 
complementary. In the usual matrix representation 
$$
a_1=\left[\begin{array}{cc} 0 & 1 \\ 0 & 0 \end{array}\right]
\otimes \left[\begin{array}{cc} 1 & 0 \\ 0 & 1 \end{array}\right]
\quad \mbox{and}\quad
a_2=\left[\begin{array}{cc} 1 & 0 \\ 0 & -1 \end{array}\right]
\otimes \left[\begin{array}{cc} 0 & 1 \\ 0 & 0 \end{array}\right],
$$
therefore
$$
\iA_1=\left\{\left[\begin{array}{rrrr}
a & 0 & b & 0\\
0 & a & 0 & b\\
c & 0 & d & 0\\
0 & c & 0 & d\\
\end{array}\right]\right\},\quad
\iA_2=\left\{\left[\begin{array}{rrrr}
a & b & 0 & 0\\
c & d & 0 & 0\\
0 & 0 & a & -b\\
0 & 0 & -c & d\\
\end{array}\right]\right\}\,.
$$

The subalgebra $\iA_1$ is generated by the F-triplet
$$
(\sigma_1 \ot \sigma_0,  \quad \sigma_2 \ot \sigma_0, \quad \sigma_3 \ot \sigma_0),
$$
and $\iA_2$ is spanned by the F-triplet
$$
(\sigma_3 \ot \sigma_1,  \quad \sigma_3 \ot \sigma_2, \quad \sigma_0 \ot \sigma_3).
$$
Observe that the standard Bell triplet (\ref{Be-tri}) is complementary 
to both $\iA_1$ and $\iA_2$. 

The parity automorphism is defined by $\Theta(a_1)= -a_1$ and $\Theta(a_2)= -a_2$.
It is induced by the unitary $\sigma_3 \ot \sigma_3$:
$$
\Theta(x)=(\sigma_3 \ot \sigma_3)x(\sigma_3 \ot \sigma_3)
$$
The operators $\sigma_i \ot \sigma_j$, $0 \le i,j\le 3$ are eigenvectors
of the parity automorphism. The fixed point algebra is linearly spanned by
$$
\sigma_0 \ot \sigma_0, \quad \sigma_1 \ot \sigma_1, \quad \sigma_2 \ot \sigma_2,
 \quad \sigma_3 \ot \sigma_3
$$
and
$$
\sigma_0 \ot \sigma_3,  \quad \sigma_1 \ot \sigma_2, \quad \sigma_2 \ot \sigma_1,
 \quad \sigma_3 \ot \sigma_0.
$$
The first group linearly spans the M-subalgebra corresponding to the Bell
basis. It follows that all Bell states are {\bf even}, that is the parity 
automorphism $\Theta$ leaves them invariant.\qed
\end{pl}

\section{Complementary decompositions}

In this section the complementary decompositions of $\iM\equiv M_4(\bbbc)$ into F- and
M-subalgebras are studied. It is well-known that decomposition into 5 M-subalgebras
is possible. (Recall that this fact is equivalent to the existence of 5 mutually
unbiased bases in a 4 dimensional space.)

The traceless dimension  $\tldim\, \iA:= \dim \left (\iA \ominus \bbbc I \right)=
\dim \iA -1$ will be used sometimes.

\begin{Thm}
Let $\iA_0$ be an F-subalgebra of $\iM$, $\iA_0'$ be its commutant, and let $\iB$ 
be a subalgebra complementary to $\iA_0$. 
\begin{itemize}
\item[(a)]
If $\iB$ is an M-subalgebra, then it is complementary to $\iA_0'$.
\item[(b)]
If $\iB$ is an F-subalgebra, then either $\tldim \left(\iA_0' \cap \iB\right)=1$ 
or $\iA_0'=\iB$.
\end{itemize}
\end{Thm}

\proof
In case (a), let $(X,Y,Z)$ the M-triplet generating $\iB$.
We can assume, that $\iA_0=M_2(\bbbc) \otimes \bbbc I$.

We can also presume, that $X,Y,Z\notin\iA'$, or otherwise
if $X=I \otimes A$, then the commutant of $X$ is generated by the matrices 
$$
V \otimes A, V \otimes I \qquad (V\in M_2(\bbbc))
$$ 
and because of the complementarity of $\iB$ and $\iA_0$, $Y=V_1 \otimes A$ holds, 
and so $Z=XY=V_1 \otimes I \in \iA_0$. This is a contradiction.

We can take the expansion
$$
X=\sum_{i=0}^3 (x_i \cdot \sigma) \ot \sigma_i,
$$
and from $X^2=I$ we conclude that $x_i \times x_j=0$ 
when  $i\ne j$ and they are different from $0$. There is an $x\in\bbbr^3$ 
such that
$$
X=(x\cdot\sigma) \otimes (\lambda \cdot \sigma)+x_0\cdot \sigma \otimes I
$$
holds, and similarly there are $y,z,y_0,z_0 \in\bbbr^3$ vectors, such that
$$
Y=(y\cdot\sigma) \otimes (\lambda \cdot \sigma)+y_0\cdot \sigma \otimes I,
$$
$$
Z=(z\cdot\sigma) \otimes (\gamma \cdot \sigma)+z_0\cdot \sigma \otimes I.
$$
{F}rom $X^2=I$ we can also conclude, that $x \bot x_0$, $y \bot y_0$ and $z \bot z_0$.

Now, $Z=XY$ is traceless and self-adjoint, so
\begin{eqnarray*}
XY&=& (x\cdot\sigma)(y\cdot \sigma)\otimes(\mu\cdot\sigma)
(\lambda\cdot\sigma)+(x\cdot\sigma)(y_0\cdot\sigma)\otimes\mu\cdot\sigma\\
&&{}+(x_0\cdot\sigma)(y\cdot\sigma)\otimes\lambda\cdot\sigma+
(x_0\cdot\sigma)(y_0\cdot\sigma)\otimes I\\
&=& -(x\times y)\cdot\sigma\otimes(\mu \times \lambda)
\cdot\sigma+\left(\<x,y\>\<\mu,\lambda\>+\<x_0,y_0\>\right)I \otimes I\\
&&{}+I\otimes (\<x,y_0\>\mu+\<x_0,y\>\lambda)\cdot\sigma\\
&&{}+\im ( (\<\mu,\lambda\>(x \times y)+(x_0\times y_0))\cdot\sigma 
\otimes I + I\otimes \<x,y\>(\mu \times \lambda)\cdot\sigma\\
&&{}+(x\times y_0)\cdot\sigma \otimes \mu\cdot\sigma + (x_0\times y)
\cdot\sigma \otimes \lambda\cdot\sigma)\\
&=&-(x\times y)\cdot\sigma\otimes(\mu \times \lambda)\cdot\sigma\\
&=&z\cdot\sigma\otimes \gamma\cdot\sigma.\\
\end{eqnarray*}
so $z_0=0$, and
\begin{eqnarray*}
ZX=Y&=&(z\cdot\sigma)(x\cdot\sigma)\otimes (\gamma\cdot\sigma)
(\mu\cdot\sigma)+(z\cdot\sigma)(x_0\cdot\sigma)\otimes I\\
&=&-(z\times x)\cdot\sigma \otimes (\gamma \times \mu)\cdot\sigma \\
&&{}+(\<z,x\>\<\gamma,\mu\>+\<z,x_0\>)I\otimes I\\
&&{}+i((z\times x_0)+\<\gamma,\mu\>(z\times x))\cdot\sigma \otimes I.\\
\end{eqnarray*}
Again, $Y$ is self-adjoint, so $y_0=0$, and similarly $x_0=0$.

(b) follows from \cite{OPSz}. \qed

Although  $\iM$ has 5 pairwise complementary M-subalgebras,
it does not have 5 pairwise complementary F-subalgebras \cite{nofive}. 
The next theorem describes the possible complementary decompositions.

\begin{Thm}
Let $\iA_k$ ($0 \le k \le 4$) be pairwise complementary subalgebras of $\iM$ 
such that all of them is an F-subalgebra or M-subalgebra. If $\ell$ is the number 
of F-subalgebras in the set $\{\iA_k \,:\, 0 \le k \le 4\}$, then 
$\ell\in\{0,2,4\}$, and all those values are actually possible.
\end{Thm}

\proof
First we give an example of $\ell=0$. The M-triplet
$$
\{\sigma_{12},\sigma_{23},\sigma_{31}\},\, 
\{\sigma_{13},\sigma_{21},\sigma_{32}\},\,
\{\sigma_{01},\sigma_{10},\sigma_{11}\},\,
\{\sigma_{02},\sigma_{20},\sigma_{22}\},\,
\{\sigma_{03},\sigma_{30},\sigma_{33}\}
$$
give the M-subalgebras. (Note that this case is about five mutually unbiased bases.)

$\ell=1$ is not possible, because if $\iA_0$ is an F-algebra, then $\{\iA_i\}_{i>0}$ 
are also complementary to $\iA_0'$, so $\tldim \left( \bigcup_{i>0} \iA_i \right) 
\le 9$, so they cannot be pairwise complementary, and this is a contradiction.

$\ell=2$  is possible. The F-triplets
$$
\{\sigma_{01},\sigma_{02},\sigma_{03}\}, \,\{\sigma_{10},\sigma_{20},\sigma_{30}\}
$$
and the M-triplets
$$
\{\sigma_{11},\sigma_{22},\sigma_{33}\},\, \{\sigma_{12},\sigma_{23},\sigma_{31}\},\,
\{\sigma_{13},\sigma_{21},\sigma_{32}\}
$$
determine the subalgebras.

${\ell=3}$ is not possible, because if $\iA_0,\iA_1,\iA_2$ are F-algebras, then 
$\{\iA_i\}_{i>2}$ are also complementary to $\iA_0'$ and $\iA_1'$. It is easy 
to see, that $\tldim \left(\bigcup_{i<3} \iA_i \cup \iA_0' \cup \iA_1'\right) 
\ge 10$, so $\tldim \left(\bigcup_{i>2} \iA_i\right)\le 5$, and they cannot 
be pairwise complementary.

${\ell=4}$  is possible. The F-triplets are
$$
\{\sigma_{01},\sigma_{02},\sigma_{03}\}, \, \{\sigma_{10},\sigma_{21},\sigma_{31}\},\,
\{\sigma_{20},\sigma_{12},\sigma_{32}\},\,\{\sigma_{30},\sigma_{13},\sigma_{23}\}
$$
and the M-triplet $\{\sigma_{11},\sigma_{22},\sigma_{33}\}$ spans the Bell subalgebra.
It was proved in \cite{OPSz} that this kind of decomposition is essentially unique,
given 4 pairwise complementary F-subalgebras the rest is always an M-subalgebra.
It follows that ${\ell=5}$ is not possible. \qed

\section{Discussion and conclusion}
The motivation for complementary subalgebras was a certain kind of state tomography
for two qubits \cite{PHSz} and a systematic study started in \cite{PDcomp}. An 
M-subalgebra (corresponding to a measurement) may give classical information and 
a (non-commutative) F-subalgebra quantum information about the total system. The
complementarity of M-subalgebras is characterized by conditional entropy, however
the single data of conditional entropy does not give the complementarity of 
F-subalgebras.

The construction of complementary subalgebras needs much research. For a 4-level 
quantum system a complete description is given in the paper. There is no F-subalgebra 
complementary to both qubits and there is essentially one M-subalgebra complementary 
to both qubits, this corresponds to the Bell basis.

The difference between $M_2(\bbbc)\ot M_2(\bbbc)$ and $M_n(\bbbc)\ot M_n(\bbbc)$
is essential. The dimensional upper bound for the number of complementary 
subalgebras (isomorphic to $M_n(\bbbc)$) is $n^2+1$. This bound is not reached
for $n=2$ \cite{nofive} but it is reached if $n>2$ is a prime \cite{Ohno}.
Some related conjecture is contained in \cite{OPSz}.

\section*{Appendix}

Let $\iA,\iB\subset M_n(\bbbc)$ be maximal Abelian subalgebras with minimal 
projections $p_1,p_2, \dots, p_n$ and $q_1,q_2,\dots, q_n$. It was expected in 
\cite{Choda} that
\begin{equation}\label{E:conj}
H(\iA|\iB)= \frac{1}{n}\sum_{i,j=1}^n \eta(\Tr (p_iq_j)).
\end{equation}
We want to analyse the case $n=2$ and show that this formula is not true.
(Unfortunately, the correction of the formula seems to be a hard problem.)

Let $\iA$ be the algebra of diagonal matrices, the algebra generated by 
$\sigma_3$ and let $\iB$ be the algebra generated by $\sin \beta \sigma_1
+ \cos \beta \sigma_3$. Then the minimal projections are
$$
p:=\frac{1}{2}(I+\sigma_3)=\left[\matrix{ 1 & 0 \cr 0 & 0 }\right], \qquad
q=\frac{1}{2}(I+(\sin \beta) \sigma_1+ (\cos \beta) \sigma_3).
$$
Then the right-hand-side of (\ref{E:conj}) is
$$
C:=\eta\Big(\frac{1+\cos \beta}{2}\Big)+\eta\Big(\frac{1-\cos \beta}{2}\Big).
$$

Let $r(t)$ be a parametrized family of minimal projections in $M_2(\bbbc)$. Then 
$$
I=\frac{1}{2} (2r(t)) +\frac{1}{2} (2I-2r(t))
$$
can be considered as a convex decomposition of the identity
in (\ref{eta}) and we have
$$
H(\iA|\iB) \ge \eta(b(t))+\eta(1-b(t))-\eta (a(t))-\eta(1-a(t))=:f(t)
$$
for all $t\in \bbbr$,
where $a(t):= \Tr (r(t)p)$ and $b:=\Tr (r(t)q)$. Choose
$$
r(t):=\frac{1}{2}(I+(\sin t) \sigma_1+ (\cos t) \sigma_3).
$$
Then $f(0)$ equals to $C$. However, it is a matter of computation
to check that $f$ is differentiable, and that if $\beta$ is such that 
neither $\sin \beta\neq 0$, nor $\cos \beta\neq 0$, then  $f'(0)\neq 
0$. Thus in general $f$ cannot have a maximum at $t=0$, and hence
(\ref{E:conj}) is not true.

\end{document}